# Ultrafast exciton dissociation and long-lived charge separation in a photovoltaic pentacene-MoS$_2$ van der Waals heterojunction

*Stephanie Bettis Homan,[1,#] Vinod K. Sangwan,[2,#] Itamar Balla,[2] Hadallia Bergeron,[2] Emily A. Weiss,[1,2*] and Mark C. Hersam[1,2,*]*

[1]Department of Chemistry, Northwestern Univeristy, Evanston, Illinois 60208, USA

[2]Department of Materials Science and Engineering, Northwestern University, Evanston, Illinois 60208, USA

[#]These authors contributed equally

*E-mails: m-hersam@northwestern.edu; e-weiss@northwestern.edu

**ABSTRACT:**

Van der Waals heterojunctions between two-dimensional (2D) layered materials and nanomaterials of different dimensions present unprecedented opportunites for novel gate-tunable optoelectronic devices. Mixed dimension p-n heterojunction diodes, such as p-type pentacene (0D) and n-type monolayer MoS$_2$ (2D), are especially interesting for photovoltaic applications where absorption cross-section and charge transfer processes can be tailored by rational selection from the vast library for 2D, organic, and inorganic photovoltaic materials. Here, for the first time, we study the kinetics of excited carriers in pentacene-MoS$_2$ p-n type-II heterojunction by transient absorption sepctroscopy. We observe the dissociation of MoS$_2$ excitons by a hole transfer to pentacene on the time scale of 6.7 ps, and a long-lived (5.1 ns) charge-separated state that is 2 to 60 times the recombination lifetime in previously reported n-n$^+$ 2D heterojunctions . By studying fractional amplitudes of the MoS$_2$ decay processes we determine a 50% hole transfer yield from MoS$_2$ to pentacene, where the remaining holes undergo ultrafast trapping due to surface defects. The ultrafast charge transfer and long-lived charge-separated state in this pentacene-MoS$_2$ Van der Waals heterojunction fulfills the requirements for high-performance photovoltaics using mixed dimensional van der Waals nanomaterials, and provide a platform for the development of future devices.

**TEXT:**

Two-dimensional (2D) van der Waals heterojunctions have emerged as a platform for unprecedented electronic function within devices[1-3] such as tunneling transistors,[4, 5] gate-tunable Schottky diodes,[6] gate-tunable p-n heterojunction diodes,[7-9] ultrafast photodetectors,[10] light-emitting diodes,[11] and photovoltaic cells[12]. While the photoinduced charge transfer that underlies optoelectronic functionality has been quantified in 2D-2D van der Waals heterojunctions,[10, 13-19] the systems studied to date have either been n-n$^+$ junctions or Schottky junctions between a 2D semiconductor and graphene. For photovoltaic applications, however, a type-II p-n heterojunction is necessary to achieve significant open circuit voltages, as has been observed for pentacene-MoS$_2$ bilayers where pentacene and MoS$_2$ act as p-type molecular donors and n-type 2D acceptors, respectively.[7] Here, for the first time, ultrafast (6.7 ps) exciton dissociation is characterized for this class of van der Waals type-II p-n heterojunctions using transient absorption spectroscopy. Unlike n-n+ heterojunctions, pentacene-MoS$_2$ p-n heterojunctions are found to possess long-lived (5.1 ns) charge-separated states that are critical for high-performance photovoltaics.

As schematically shown in **Figure 1a**, a monolayer film of n-type $MoS_2$ forms a van der Waals p-n heterojunction with p-type pentacene. Experimentally, a large-area monolayer $MoS_2$ film was grown directly on quartz wafers by chemical vapor deposition (CVD) modifying the recipe reported previously (Supporting Information Section S1).[20, 21] This direct growth method yields a homogeneous surface (without further processing) for subsequent deposition of pentacene.[22] The monolayer thickness and stoichiometry of $MoS_2$ was confirmed by atomic force microscopy (AFM), Raman microscopy, and X-ray photoelectron spectroscopy (Supporting Information Sections S1–S4). A 30 nm thick continuous pentacene film was subsequently grown by thermal evaporation in a $N_2$ glove box to minimize ambient-induced degradation (Supporting Information Section S3).[7] AFM images show that pentacene crystallizes into approximately 500×500 $nm^2$ grains on $MoS_2$, even though the first few layers of pentacene do not form ordered assemblies at the $MoS_2$ interface[22] (**Figure 1b**).

**Figure 1c** shows the ground-state absorption spectra of the heterojunction and the isolated components of the junction. The pentacene spectrum has peaks at 584 nm, 630 nm, and 670 nm, the sum of absorptions from H- and J-aggregates, charge transfer states, and associated vibrational sidebands.[23] The $MoS_2$ spectrum includes the B and A excitons of monolayer $MoS_2$ at 605 nm and 648 nm, respectively. These peaks shift to lower energy upon deposition of pentacene due to the spontaneous flow of electrons from $MoS_2$ to pentacene and the formation of a depletion region. The decrease in the bandgap of monolayer $MoS_2$ upon removal of free electrons has also been observed by photoluminescence (PL)[24] and scanning tunneling spectroscopy.[25] The spectrum of $WS_2$ within a van der Waals heterojunction with graphene has also exhibited a bathochromic shift of similar magnitude (13 meV) due to the formation of a depletion region near the Schottky contact.[17] Van der Waals heterojunction between $MoSe_2$ and $WS_2$ has also shown a bathochromic shift of 20 meV.[26] The PL spectrum of the MoS$_2$-only monolayer upon excitation at 532 nm (**Figure 1d**) is dominated by emission from the A-exciton at 654 nm, and the PL spectrum of the pentacene-only film has broad peaks at 585 nm and 680 nm.[20, 27] Within the heterojunction, the PL from the MoS$_2$ A-exciton is quenched by 83% (Supporting Information Section S4). The quenching of the pentacene PL in the heterojunction cannot be quantified due to variations in pentacene thickness within the laser spot (~1 μm$^2$) (**Figure 1b**).

**Figure 2** shows the two-dimensional (2D) transient absorption (TA) spectra of the heterojunction, and of the isolated MoS$_2$ and pentacene films, after photoexcitation at 535 nm, while **Figure 3a** shows these spectra at a specific pump-probe delay of 500 fs. The spectrum of the MoS$_2$-only film is dominated by the bleaches of the B-exciton at 610 nm and the A-exciton at 650 nm (**Figures 2a, 3a**). These bleaches are shifted to lower energy by 6 meV and 17 meV, respectively, from their positions in the ground-state absorption spectrum, due to photoinduced bandgap renormalization, which is the net result of competitive contributions from electronic gap shrinking (red shift) and reduction of the exciton binding energy (blue shift) upon excitation at the band-edge.[28] At later times, the MoS$_2$-only film exhibits a hypsochromic (blue) shift of the bleach features (seen in **Figure 2a** and quantified in Supporting Information Figure S10) that is likely attributable to exciton-phonon scattering.[29, 30] The initial amplitude of the MoS$_2$ signal depends linearly on pump fluence and the dynamics are independent of excitation power, which rules out biexciton formation and hot phonon effects in this system (Supporting Information Figure S11).[15, 31] The pentacene-only film has two bleach features corresponding to its ground state absorptions at 586 nm and 683 nm (**Figure 1b**) that do not decay on the timescale of this measurement (**Figure 2b**). The absence of a bleach feature at 635 nm and the overall weaker intensity of the pentacene TA signal relative to the MoS$_2$ signal, despite the greater ground-state absorbance of pentacene, have been previously attributed to overlap of these bleach signals with absorptions of the pentacene triplet excited state, which forms by singlet fission within 80 fs of photoexcitation.[32, 33]

Analysis of the kinetic trace extracted at 612 nm from the B-exciton feature of the MoS$_2$-only film (**Figure 3b**) allows us to list the mechanisms for exciton decay in this material and their time constants, as seen in **Table 1** and **Figure 4a**. The fastest component measured ($\tau_1$ = 670 fs) has been reported previously for CVD-grown MoS$_2$[29, 34] and exfoliated MoS$_2$ flakes,[35] and is attributed to quenching of the MoS$_2$ exciton by carrier trapping. The second component, $\tau_3$ = 15.8 ps, is associated with the exciton-phonon scattering process mentioned above.[29, 30] The third component, $\tau_4$ = 431 ps, is consistent with previously reported time constants for radiative recombination of the exciton (250 – 850 ps).[29, 30, 36-38] A fractional amplitude of 0.18 for the radiative recombination component is consistent with previous TA studies,[30] but not with the reports of low photoluminescence quantum yield of the MoS$_2$ monolayer of 0.4%.[39] We therefore suspect that electron trapping, which typically occurs in hundreds of picoseconds in transition metal dichalcogenides, is convolved with the radiative decay component in this kinetic trace.[35]

The main features in the TA spectrum of the heterojunction are again the two bleaches of the B-exciton and A-exciton of MoS$_2$ (at 612 nm and 666 nm, respectively) shifted further to lower energy from those in the MoS$_2$-only TA spectrum, consistent with the ground state absorption spectrum (**Figure 1c**). The hypsochromic (blue) shift of the MoS$_2$ ground state bleach is no longer observed in the 2D TA spectra (**Figure 2b**), and instead the spectrum appears to shift to lower energy at later times due to contributions from the directly excited pentacene bleach. We monitored the dynamics of the MoS$_2$-pentacene heterojunction at 612 nm since the kinetic traces only contain signal from the MoS$_2$ B-exciton at this wavelength. Importantly, this wavelength possesses no contribution from the pentacene excited state (see the orange dots in **Figure 3b**) or from the pentacene radical cation, as confirmed by spectroelectrochemistry (Supporting Information Section S6). The dynamics of the $MoS_2$ exciton in the heterojunction differ from those in the $MoS_2$-only film, probed at the same wavelength, in two ways: (i) the exciton-phonon scattering component ($\tau_3$) is replaced with a 6.7 ps decay ($\tau_2$); (ii) an additional decay component of 5.1 ns ($\tau_5$) is present, Table 1. Given the quenching of the $MoS_2$ PL in the heterojunction, and the absence of the pentacene excited state in the TA spectra (which would form if energy transfer were occurring), we assign the 6.7 ps component to hole transfer from photoexcited n-type $MoS_2$ to p-type pentacene and the 5.1 ns component to the recombination of this transferred hole with an excess electron in $MoS_2$ (**Figure 4b**). The hole transfer process in the pentacene-$MoS_2$ heterojunction is at least a factor of ten slower than that within 2D-2D n-n$^+$ heterojunctions[13, 14, 18] (<400 fs). We attribute this difference in kinetics to the weaker inter-layer coupling at the structurally disordered interface between $MoS_2$ and pentacene.[22] This charge recombination lifetime of ~5 ns is shorter than that in polymer/fullerene bulk heterojunction solar cells (20 ns-10 μs),[40] but it is 2 to 60 times longer than the lifetime of indirect excitons in planar 2D-2D $MoSe_2$-$WS_2$ (80 ps),[15] $MoS_2$-$MoSe_2$ (240 ps)[14] and $MoSe_2$-$WSe_2$ (1.8 ns)[16] heterojunctions.

Analysis of the fractional amplitudes of the decay components in the kinetic trace of the heterojunction allows us to estimate an overall hole transfer yield of ~50%. The fast trapping of electrons and holes[35] to surface defects ($\tau_1$ = 670 fs) accounts for ~48% of total exciton decay in both the $MoS_2$-only and heterojunction samples. It has been shown previously that the amplitudes of the A-exciton and B-exciton bleaches in the $MoS_2$-only TA spectrum do not depend on which exciton is directly excited (i.e., which sub-band the hole occupies), and the total degeneracy of both the conduction and valence bands is two, considering both excitons.[41] We therefore estimate that the electron and hole contribute equal amplitudes to the ground state bleach of $MoS_2$. Equal partitioning of the remaining ~50% of the bleach amplitude between electron and hole is achieved by assigning $\tau_5$ and $\tau_4$ to electron relaxation by charge recombination with the transferred hole (5 ns, 15%) and electron trapping (431 ps, 9%), respectively. Radiative recombination is a negligible contribution to bleach dynamics. With these assignments, all of the hole dynamics after the fast ($\tau_1$) trapping process are accounted for by hole transfer ($\tau_2$), implying that the overall yield of hole transfer in the heterojunction is $100\% - 48\% \cong 50\%$.

These results indicate that the rate of hole transfer in the pentacene-$MoS_2$ p-n heterojunction is fast enough to out-compete all hole relaxation processes except sub-picosecond carrier trapping. While charge transfer in n-n$^+$ 2D-2D heterojunctions competes more favorably with this ultrafast decay process, and therefore potentially leads to a higher yield of exciton dissociation than the 50% we observe for the pentacene-$MoS_2$ p-n heterojunction, the same strong inter-layer coupling in the 2D-2D systems that enables fast charge separation also provides pathways for fast charge recombination. This faster recombination ultimately limits the ability of charge carriers to diffuse or drift away from the dissociation site, thereby limiting charge collection efficiency in a photovoltaic cell. Furthermore, the pentacene-$MoS_2$ p-n heterojunction has a larger theoretical maximum open circuit voltage (~1.1 V from free carrier band-offset) than n-n$^+$ 2D-2D heterojunctions (0.2–0.4 V), further confirming its suitability for photovoltaic applications[13, 14] With that said, a limitation of the pentacene-$MoS_2$ p-n heterojunction is the sub-picosecond singlet fission process in pentacene that creates two low energy triplet states and thereby eliminates the contribution of excitons in pentacene to the photocurrent. In light of this discussion, replacement of pentacene with a p-type organic molecule or polymer with similar energy level alignment, but a longer-lived singlet excited state, appears particularly promising. Overall, by elucidating the dynamics in a prototypical type-II p-n heterojunction, this study will inform ongoing efforts to employ molecular donors and 2D acceptors in van der Waals heterostructure photovoltaics.

## ASSOCIATED CONTENT

**Supporting Information**:

Additional details on experimental methods, electrical characterization, simulations, and TA data accompanies this paper and is available free of charge via the Internet at http://pubs.acs.org.

## AUTHOR INFORMATION

**Corresponding authors:**

**\***E-mail: m-hersam@northwestern.edu; e-weiss@northwestern.edu

## AUTHOR CONTRIBUTIONS:

M.C.H., E.A.W., S.B.H., and V.K.S conceived the idea and designed the experiments. S.B.H. conducted the transient absorption (TA) experiments. S.B.H and E.A.W analyzed the TA data. I.B. and H.B. conducted growth of $MoS_2$. V.K.S fabricated the $MoS_2$-pentacene heterojunctions. V.K.S., I.B, and H.B performed materials characterization (PL/Raman/AFM/XPS). All authors wrote the manuscript and discussed the results at all stages. #These authors contributed equally

## NOTES:

**Competing financial interests**: The authors declare no competing financial interests.

**ACKNOWLEDGMENTS:**

This research was supported by the Materials Research Science and Engineering Center (MRSEC) of Northwestern University (NSF DMR-1121262) and the 2-DARE program (NSF EFRI-143510). CVD growth of $MoS_2$ was supported by the National Institute of Standards and Technology (NIST CHiMaD 70NANB14H012). The Raman instrumentation was funded by the Argonne−Northwestern Solar Energy Research (ANSER) Energy Frontier Research Center (DOE DE-SC0001059). H.B further acknowledges support from the NSERC Postgraduate Scholarship-Doctoral Program. The authors thank Riccardo Turrisi for assistance in purification of pentacene, and Dr. Alexander Nepomnyashchii for spectroelectrochemical measurements of the pentacene radical cation. The authors also thank Dr. Sarah L. Howell and Prof. Lincoln J. Lauhon for valuable discussions.

**FIGURES:**

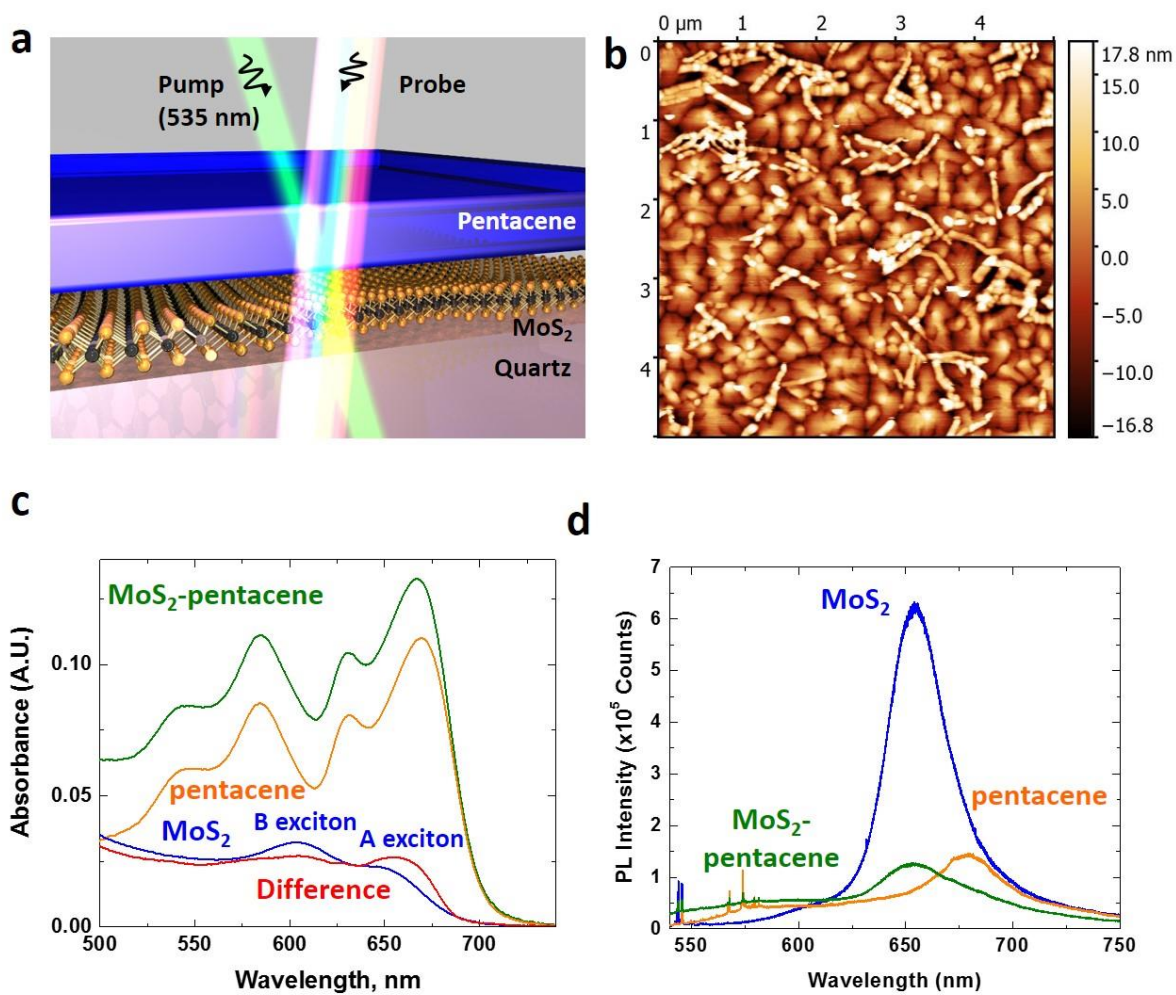

**Figure 1. Structure and steady-state characterization of the MoS$_2$-pentacene heterojunction.** (a) Schematic of the monolayer MoS$_2$-pentacene van der Waals heterojunction probed by transient absorption spectroscopy. (b) Atomic force microscope topographical image of a 30 nm thick pentacene film deposited on monolayer MoS$_2$. The maximum thickness variation of the pentacene film on MoS$_2$ is 11 nm (valley to peak). (c) Ground state absorption spectra of a MoS$_2$-pentacene heterojunction (green), a MoS$_2$-only monolayer film (scaled by 0.7 to match the difference

spectrum, blue), a 30 nm thick pentacene-only film (scaled by 0.7 to account for the difference in film size, orange), and the difference between the spectrum of the MoS$_2$-pentacene heterojunction and the scaled pentacene-only film (red). The MoS$_2$ A-exciton and B-exciton peaks are at lower energy in the heterojunction (608 nm and 651 nm, respectively) than in the MoS$_2$-only film (605 nm and 648 nm). The ground-state absorption spectra were collected using an integrating sphere. (d) Photoluminescence microscopy spectra of the MoS$_2$-only monolayer film (blue), the pentacene-only film (orange), and the MoS$_2$-pentacene heterojunction (green). The excitation wavelength was 532 nm. The sharp peaks at 575 nm arise from Raman scattering in pentacene, and the peaks below 550 nm arise from Raman scattering in the MoS$_2$ and quartz substrate (Supporting Information Section S4).

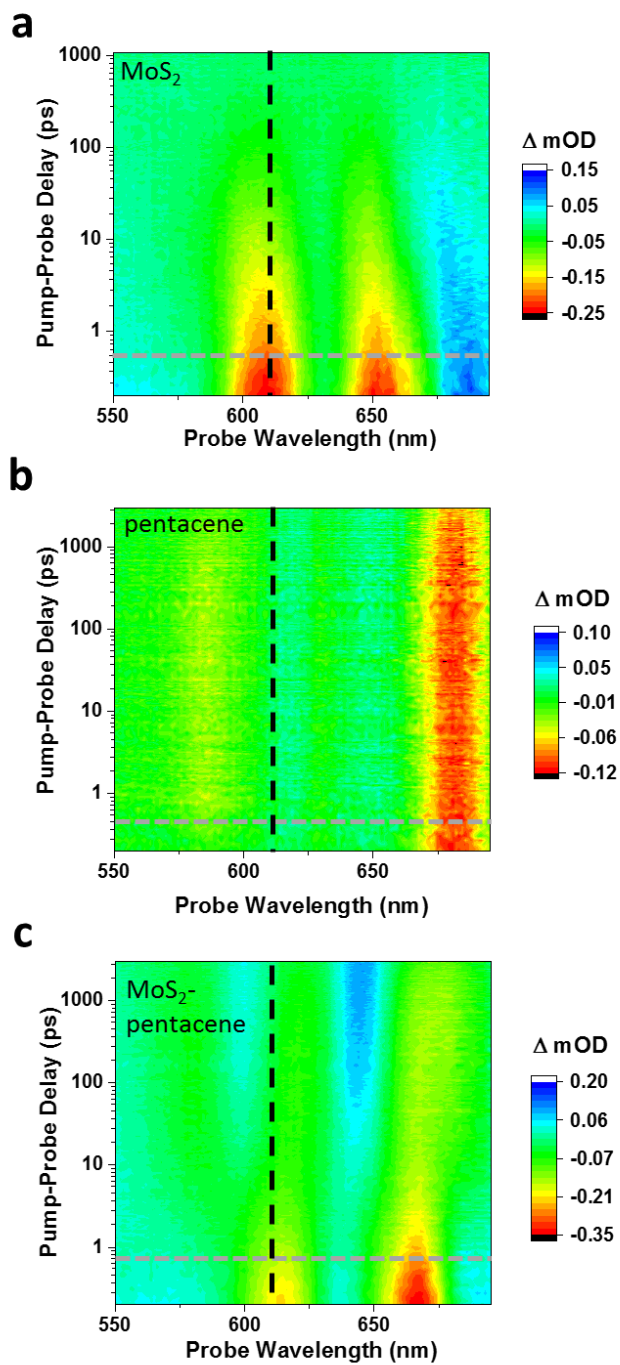

**Figure 2.** Two-dimensional transient absorption spectra. Transient absorption spectra at a range of pump-probe delay times for the (a) MoS$_2$-only film, (b) pentacene-only film, and (c) MoS$_2$-pentacene heterojunction upon excitation at 535 nm with a fluence of 4 μJ/cm$^2$. The dashed lines

indicate a pump-probe delay time of 500 fs (gray) and the probe wavelength of 612 nm (black) shown in Figure 3. All TA spectra are averaged over six different $4.3\times10^5$-$\mu m^2$ spots on each film. The $MoS_2$-pentacene sample was prepared by depositing pentacene on the same film used for the $MoS_2$-only sample, and TA experiments were performed at approximately the same locations before and after pentacene deposition.

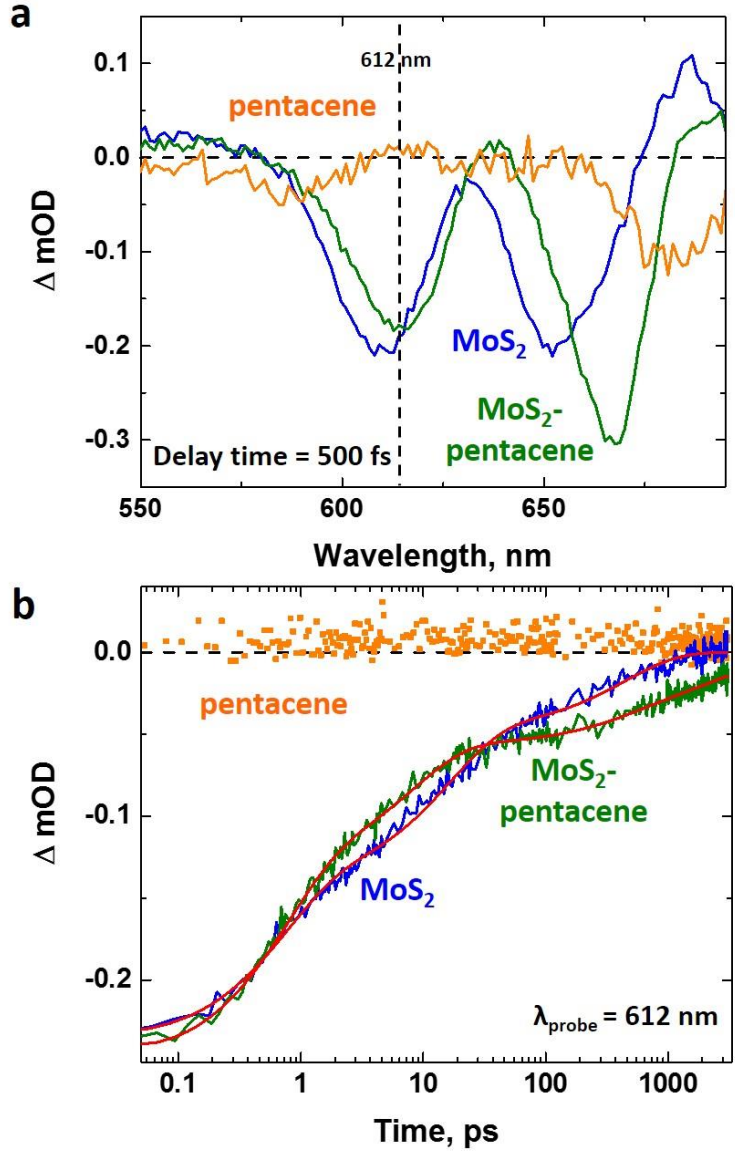

**Figure 3.** Transient absorption spectroscopy of the MoS$_2$-pentacene heterojunction. (a) Transient absorption spectra at a single pump-probe delay time of 500 fs after excitation at 535 nm for the monolayer MoS$_2$-only film (blue), the 30 nm thick pentacene-only film (orange), and the MoS$_2$–pentacene heterojunction (green). Excitation at 535 nm excites both pentacene (by 72%) and MoS$_2$ (by 28%) with a fluence of 4 µJ/cm$^2$ that generates an exciton density of 6.55 × 10$^{10}$ cm$^{-2}$ within monolayer MoS$_2$. (b) Kinetic traces extracted at the B-exciton peak (612 nm), from the averaged

TA spectra shown in part (a). The amplitude of the MoS$_2$–pentacene trace is scaled by a factor of 0.89 in order to align the signals at time-zero of the experiment. The red lines are fits of the kinetic data to a sum of exponential functions convoluted with an instrument response (IRF = 150 fs), with the parameters listed in **Table 1**. Kinetic traces and fitting parameters for each individual spot for each sample are provided in Supporting Information Section S5. Despite being excited at the pump wavelength (535 nm), the pentacene signal does not evolve in time at the probe wavelength of 612 nm.

**Table 1.** Time constants for decay of the B-exciton (monitored at 612 nm) within $MoS_2$-only and $MoS_2$-pentacene heterojunction films, after excitation at 535 nm.[a]

| | $\tau_1$ ($A_1$) <br> *carrier trapping* | $\tau_2$ ($A_2$) <br> *$h^+$ transfer* | $\tau_3$ ($A_3$) <br> *exciton-phonon scattering* | $\tau_4$ ($A_4$) <br> *radiative recombination and $e^-$ trapping* | $\tau_5$ ($A_5$) <br> *charge recombination* |
|---|---|---|---|---|---|
| $MoS_2$-only film | 670 ± 20 fs <br> (0.47) | -- | 15.8 ± 0.6 ps <br> (0.35) | 431 ± 20 ps <br> (0.18) | -- |
| $MoS_2$-pentacene junction | 670 fs[b] <br> (0.48) | **6.65 ± 0.34 ps** <br> **(0.28)** | -- | 431 ps[b] <br> (0.09) | **5.13 ± 0.44 ns** <br> **(0.15)** |

[a]Each lifetime is an average of measurements at 6 locations on two $MoS_2$-only films and one heterojunction film. The quantities in parentheses are the fractional amplitudes of each component. [b]The lifetimes in cells shaded in gray are fixed to the lifetimes measured for the $MoS_2$-only film, so no uncertainty is associated with these components in the fit.

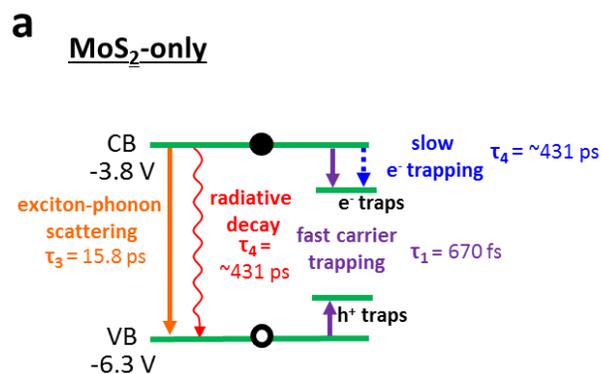

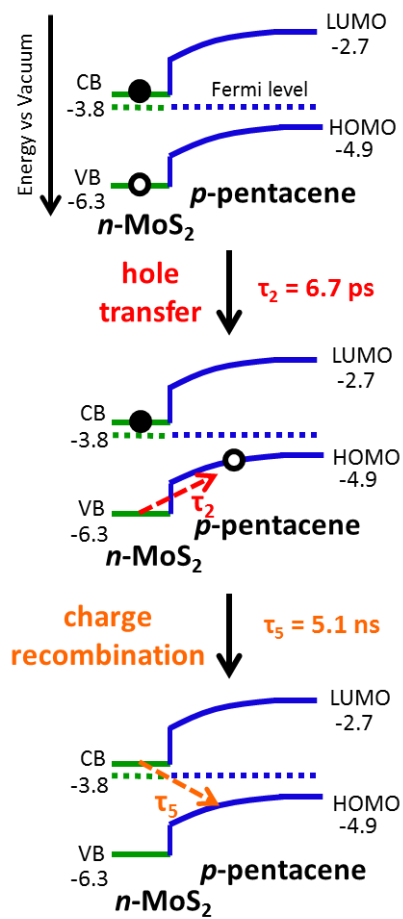

**Figure 4**. Schematic of exciton dynamics in the $MoS_2$-only film and $MoS_2$-pentacene heterojunction. (a) Summary of relaxation pathways for the monolayer $MoS_2$ exciton in the absence of pentacene, based on the dynamics of the bleach of the B-exciton feature in the TA spectrum. (b) Mechanism of hole transfer ($\tau_2 = 6.7$ ps) and charge recombination ($\tau_2 = 5.1$ ns) in the MoS$_2$-pentacene p-n heterojunction. The depletion region is shown by band-bending of pentacene at the heterojunction (band offset 1.1 V). The effect of band bending is negligible in atomically thin MoS$_2$. The MoS$_2$ valence band (VB) and conduction band (CB) energies are from references [13] and [42], and the pentacene highest occupied molecular orbital (HOMO) and lowest unoccupied molecular orbital (LUMO) potentials are from reference [43]. The electrons and holes shown in MoS$_2$ form an exciton with binding energy of ~0.5 eV.